\documentclass[preprint2]{aastex}

\newcommand{\kepler}{\textit{Kepler}}
\newcommand{\prat}{\mathcal{P}}

\shorttitle{Missing short-period planet pairs}
\shortauthors{Steffen and Farr}

\begin{document}


\title{A lack of short-period multiplanet systems with close-proximity pairs and the curious case of Kepler 42}


\author{Jason H. Steffen\altaffilmark{1,2,3} and Will M. Farr\altaffilmark{1,2}}


\altaffiltext{1}{Department of Physics and Astronomy, Northwestern University, Evanston, IL 60208}
\altaffiltext{2}{CIERA Fellow}
\altaffiltext{3}{jasonsteffen@northwestern.edu}


\begin{abstract}
Many \kepler\ multiplanet systems have planet pairs near low-order, mean-motion resonances.  In addition, many \kepler\ multiplanet systems have planets with orbital periods less than a few days.  With the exception of Kepler-42, however, there are no examples of systems with both short orbital periods and nearby companion planets while our statistical analysis predicts $\sim 17$ such pairs.  For orbital periods of the inner planet that are less than three days, the minimum period ratio of adjacent planet pairs follows the rough constraint $\prat \equiv P_2/P_1 \gtrsim 2.3 (P_1/\textrm{day})^{-2/3}$.  This absence is not due to a lack of planets with short orbital periods.  We also show a statistically significant excess of small, single candidate systems with orbital periods below 3 days over the number of multiple candidate systems with similar periods---perhaps a small-planet counterpart to the hot Jupiters.
\end{abstract}





\section{Introduction}

Since it's launch in 2009, the \kepler\ mission has provided nearly 3000 planet candidates with several hundred target stars showing multiple planet signatures \citep{Borucki:2010,Steffen:2010,Batalha:2013}.  The set of multiplanet systems are notably valuable for improving our understanding of the processes of planet formation and dynamical evolution.  For example, we can learn about the architectures of the planetary systems, the size distribution of planets within systems, and the mutual inclinations of the orbits (e.g., \citet{Lissauer:2011b,Fabrycky:2012b,Ciardi:2013,Fang:2012,Kane:2012,Ford:2011a}).  In addition, the \kepler\ planet sample is useful for making comparisons of planetary systems with different numbers of observed planets or ``multiplicity'' \citep{Latham:2011,Steffen:2012b,Steffen:2013b}.

An example of a striking difference between planetary systems with varying multiplicities is the hot Jupiter population.  The relative isolation of hot Jupiter planets compared with planets in other systems is a strong indicator of differences in their dynamical histories.  This unique architecture is seen in RV surveys, in photometric data, and in searches for additional companions using transit timing variations \citep{Agol:2005,Holman:2005,Steffen:2005,Wright:2009,Latham:2011,Steffen:2012b,Dawson:2013}.  Planets near MMR are valuable because they often indicate a history of quiescent orbital migration of planets in a system through some interaction with a gas or planetesimal disk \citep{Fernandez:1984,Malhotra:1993,Thommes:2005,Zhou:2005}.  Thus, by studying the observed orbital architectures of systems we gain insights into the histories of planetary systems.

For the inner regions of planetary systems the \kepler\ data are the premier source of information.  Of particular value is the fact that the planets in \kepler\ systems are transiting, which allows high precision measurements of their orbital phase at the time of transit (relative errors of $10^{-5}$ are common), which in turn enables effective probes of the planet-planet interactions in the system through TTVs (e.g., \citet{Holman:2010,Lissauer:2011a}).  Precise orbital periods also enable the detection of small features in the distributions of orbital periods or period ratios such as gaps or excesses of planet pairs near MMR \citep{Fabrycky:2012b,Steffen:2013b}.  Such features can indicate subtle dynamical effects like tides or differential migration \citep{Lithwick:2012a,Batygin:2013}.

A recent study by \citet{Steffen:2013b} indicated a dependence of system architecture on planet multiplicity---a difference in the proximity of planet pairs to certain MMRs was seen between ``high multiplicity'' systems with four or more planets and systems with only three or two planets.  Systems with fewer planets have more planet pairs near the 2:1 and just interior to the 3:1 MMR.  Here, we look for any dependence of the period ratios of adjacent planet pairs on the physical size of the system.  It is important to note that many relevant quantities in celestial mechanics (e.g., the Hill sphere or MMR) scale with the semi-major axis of the planet orbit.  Thus, when considering point-mass objects interacting via Newtonian gravity, it is only the period ratios that determine the dynamics of the system and not the absolute scale of the orbit.  Therefore, many of the ``tightly packed'' \kepler\ systems are, under these circumstances, quite similar to the inner part of our solar system---which has similar period ratios.

However, when physical processes enter other than pure Newtonian gravity, such as tides, disk interactions, winds, magnetic fields, or General Relativity, then the scale invariance is broken.  Consequently, the orbital architectures of the systems change only as these processes become important.  With that in mind, we look specifically at the architectures of \kepler\ planetary systems as a function of the system size (using the inner planet period as a proxy for ``size'').  If some feature emerges, it indicates that additional physics dynamically affects the either the formation or the subsequent dynamical evolution of the system.

\section{Multiplanet System Architectures}
\label{multiarch}

All data used for our analysis come from the NASA Exoplanet Archive catalog of Kepler Objects of Interest (KOIs) retrieved on June 3, 2013.  We display results in our figures using primarily the KOI catalog from Quarter 1 through Quarter 8 (Q1-Q8).  Qualitatively similar results come from an identical analysis of the catalogs through Q6 \citep{Batalha:2013} and the most recent (and least vetted) catalog through Q12.  For this study we select all KOIs that show multiple planet candidates that are not labeled as ``FALSE POSITIVE''.  We restrict our study to those with planet sizes less than 20 Earth radii (this cut eliminates 7 systems) and by requiring that the period ratio of adjacent pairs be greater than 1.1 (this cut retains Kepler-36 \citet{Carter:2012} near the 7:6 MMR but removes systems that, if orbiting a single star, would almost certainly be unstable such as KOI-284 \citep{Lissauer:2011a}).

\begin{figure}[!ht]
\includegraphics[width=0.45\textwidth]{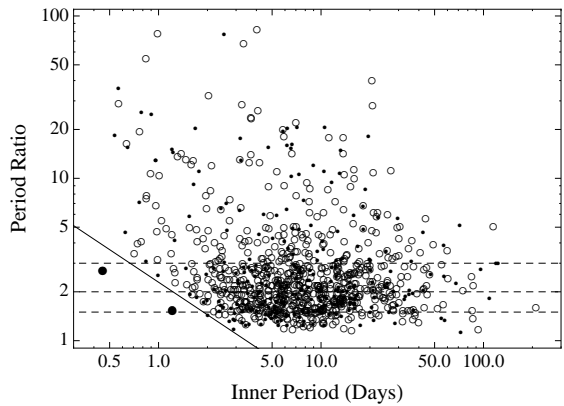}
\caption{Scatter plot of period ratio for adjacent planets as a function of the orbital period of the inner planet of that pair.  Horizontal lines indicate the 3:2, 2:1, and 3:1 MMRs.  The diagonal line is given by Equation \ref{boundaryfunc}.  Open circles are planet pairs given in the Q8 catalog, large filled circles correspond to the Kepler 42 system, and small dots correspond to either new systems or systems with new planets seen in the Q12 catalog (206 total systems).\label{pratios}}
\end{figure}

Figure \ref{pratios} shows the period ratio of each adjacent planet pair as a function of the period of the inner planet in that pair using both the Q8 catalog and the Q12 catalog.  The Q12 catalog introduces 206 new systems or systems with new planets\footnote{There are a number of points from the Q12 catalog that correspond to 49 systems with existing planet pairs in the Q8 catalog, but where a new planet candidate was found.  Since new, intermediate planets in a system would destroy the existing period ratios in that system, we chose to redisplay the entire system in this figure instead of only the new planet pairs.  Redundant pairs are those where the dot in centered in an open circle.  There is only one such point with an inner period less than one day.}.  We note that the number of observed planet pairs interior to the 3:1 MMR declines as the orbital period of the inner planet falls below a few days.  A hint of this possibility was noted in \citet{Steffen:2013b} when comparing the smallest and largest planetary systems of differing multiplicities.

The solid diagonal line in Figure \ref{pratios} is a boundary above which there are examples of planet pairs but below which no planet pairs are found except for the three planet system Kepler-42 (shown as solid circles).  This diagonal line corresponds to the curve
\begin{equation}
\label{boundaryfunc}
\prat \equiv \frac{P_2}{P_1} = 2.3 \left( \frac{P_1}{\textrm{day}} \right)^{-2/3}
\end{equation}
where $P_{1}$ and $P_2$ are the is the periods of the inner and outer planets.  It is likely that the primary quantity of interest is the distances to the planets.  Thus, two other potentially interesting representations of the curve are given by
\begin{equation}
\biggl( \frac{a_2}{\textrm{AU}} \biggr) = 0.034 \left( \frac{M_\star}{M_\odot} \right)^{1/3} \left( \frac{P_1}{\textrm{day}} \right)^{2/9}
\end{equation}
and
\begin{equation}
\biggl( \frac{a_2}{\textrm{AU}} \biggr) = 0.13 \left( \frac{M_\star}{M_\odot} \right)^{2/9} \biggl( \frac{a_1}{\textrm{AU}} \biggr)^{1/3}
\end{equation}
where $a_i$ are the semimajor axes of the respective planet orbits and $M_\star$ is the mass of the host star.  These equations are only useful for inner planet periods below $\sim 3$ days.  For larger orbital periods the formula yields unstable systems and eventually gives a (nonsense) period ratio less than unity.  We note that the exponent in this curve was assigned \textit{ad hoc} and we recognize that a viable theory need not produce this explicit functional form.

\begin{figure}[!ht]

\includegraphics[width=0.45\textwidth]{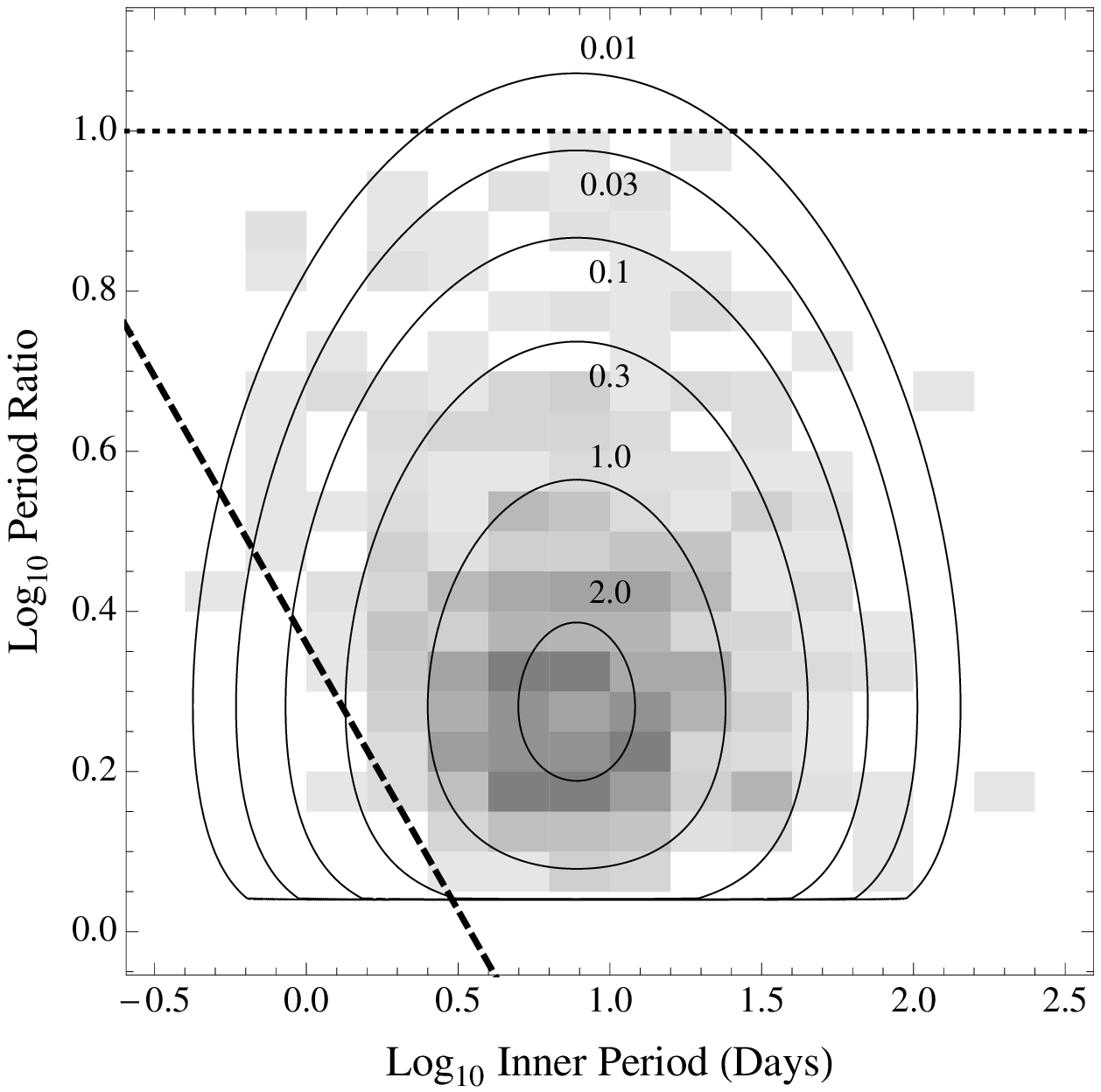}
\vskip0.3in
\includegraphics[width=0.45\textwidth]{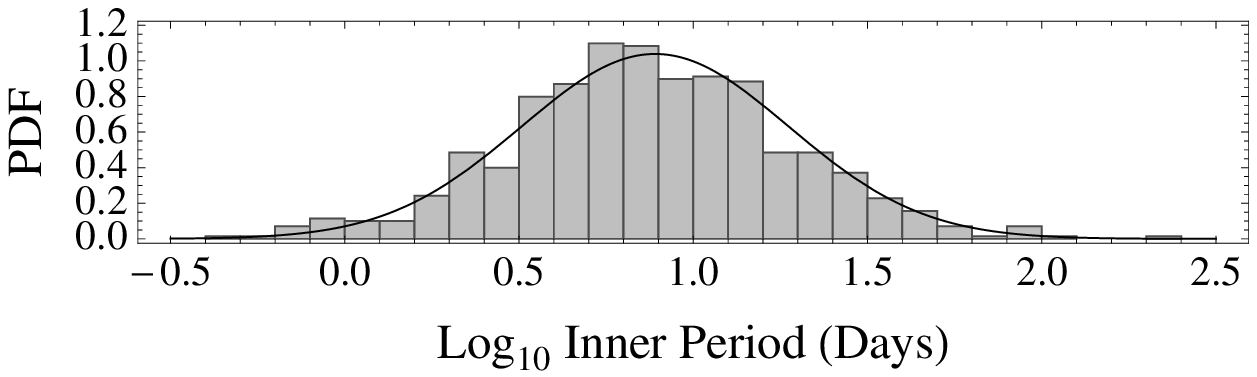}
\vskip0.3in
\includegraphics[width=0.45\textwidth]{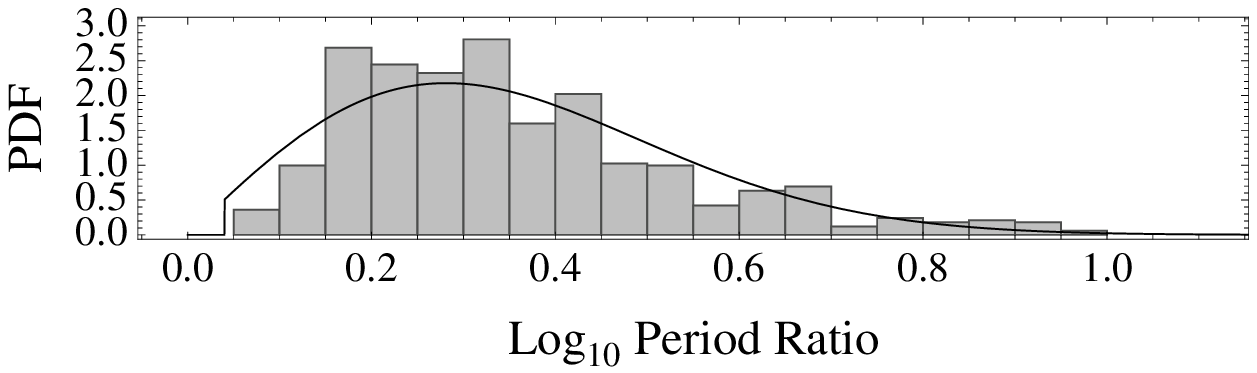}
\caption{Upper Panel: Contours of the fitted joint probability density function using a normal distribution in inner planet period and a truncated Rayleigh distribution for the period ratios.  The diagonal dashed line is Equation \ref{boundaryfunc}.  Data below the horizontal dotted line were used in the analysis.  From this fitted distribution we estimate that $\sim 17$ planet pairs should exist below the dashed line.  There are two such systems observed.  Middle Panel: The projected distribution in inner planet orbital periods along with the model fit.  Lower Panel: shows The projected period ratio distribution along with the model fit.\label{pratioscontour}}
\end{figure}

To estimate the significance of the missing systems, we fit a joint probability density function (PDF) to all data with period ratios less than 10.  The PDF is log-normal in inner planet period and is Rayleigh distributed in period ratio truncated with a lower bound of 1.1 \citep{Steffen:2013b}.  The results of this maximum likelihood fit (shown in Figure \ref{pratioscontour}) are a mean and standard deviation of 0.89 and 0.38 respectively (in log days), and a Rayleigh parameter of 0.28 (in log period ratio).

By integrating this PDF over a given region we can predict the number of expected planet pairs.  Here, the overall probability of a system being below Equation \ref{boundaryfunc} is $0.026 \pm 0.0077$.  With the 663 period ratios used in the analysis, we then expect $17.5 \pm 5.1$ pairs to lie below the boundary.  Only two are observed---indicating a statistically significant dearth of planets with a formal $p$-value of $1.2\times 10^{-3}$.  (Repeating the analysis without imposing any cuts on the data prior to fitting the PDF, reduces the number of expected planet pairs slightly to 16.3.)  Of 470 systems in the Q8 catalog, comprising 701 total planet pairs, only Kepler-42 violates the above relationship.  This trend continues using the current Q12 catalog of 590 systems and 845 planet pairs.  By way of comparison, the symmetric region in the lower right of Figure \ref{pratioscontour} contains 20 points from the Q8 catalog---consistent with the estimate from the joint PDF.


\section{Comparison with Single Planet Systems}
\label{singleplanets}

A reasonable hypothesis regarding the cause of missing planet pairs is that few planets can survive at such short orbital periods, so only an exceptional few would be in multiplanet systems.  A good comparison group is then the single-candidate systems.  If a fair number of short-orbit, single-planet systems exist, then this hypothesis would be unjustified.  To investigate this possibility, we select KOIs from the same catalog where only a single periodic transit signature is seen.  There is a known excess of hot Jupiters (perhaps 100 such systems), however these planets do not readily compare to short-period, multiplanet systems since no hot Jupiter has a known, nearby companion \citep{Steffen:2012b,Wright:2009}.  Thus, we restrict our analysis to those with estimated radii smaller than 5 Earth radii (though the results are not sensitive to this choice since the hot Jupiters only a small fraction of the total number of systems in this period range).  Our selection criteria yields 1580 single-candidate systems.

\begin{figure}[!ht]
\includegraphics[width=0.45\textwidth]{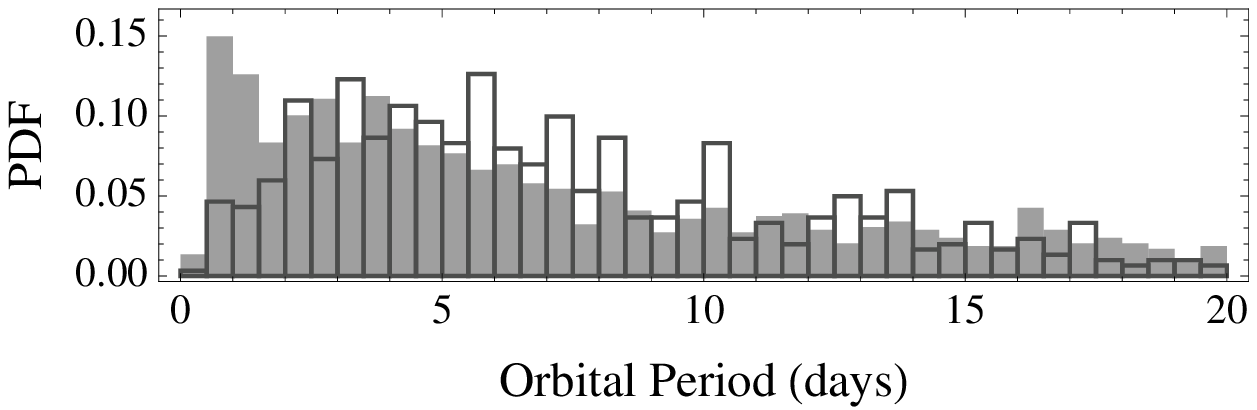}
\vskip0.3in
\includegraphics[width=0.45\textwidth]{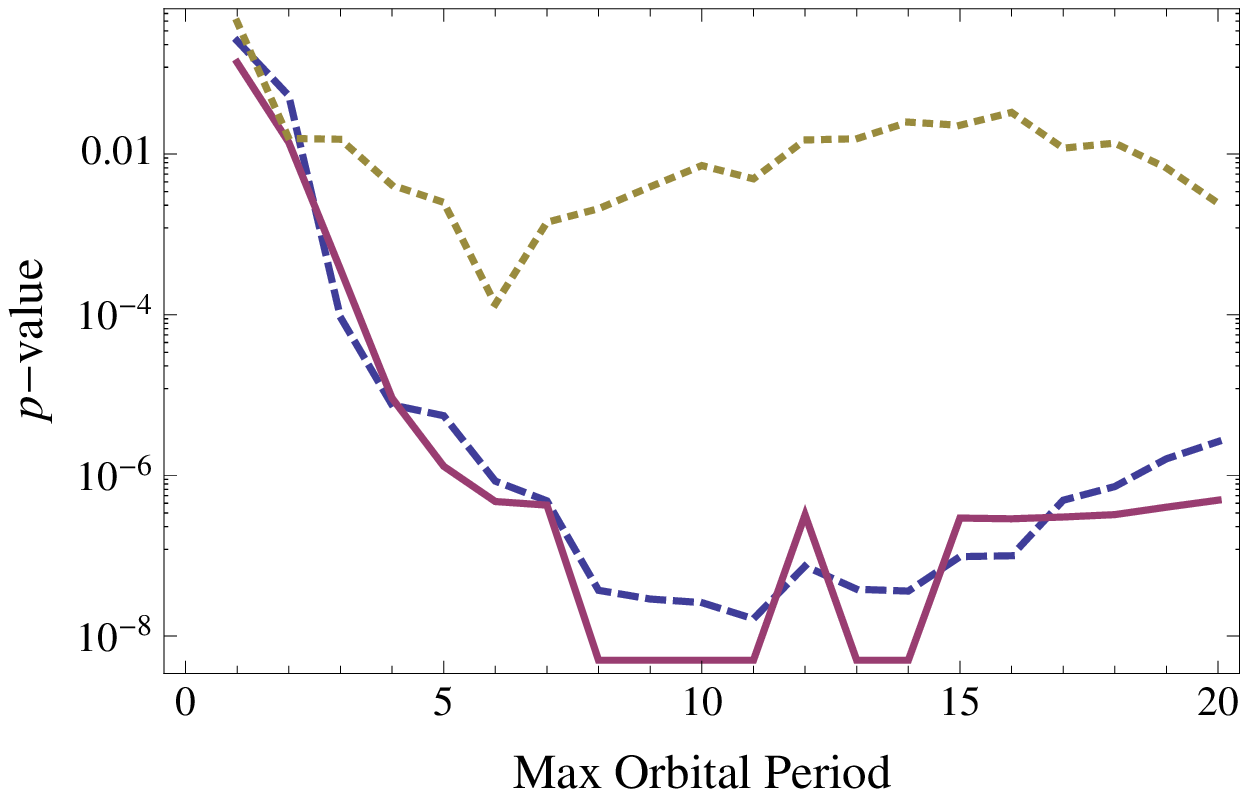}
\caption{Upper Panel: Probability density of the orbital periods for systems with single planets (shaded) and multiple planets (outline).  Hot Jupiters (planets with radii larger than 5 $R_\oplus$) are excluded for the single planet systems.  Lower Panel: Results of the KS (blue dashed) and AD (red solid) tests comparing the orbital period distribution for multiplanet and single planet systems (excluding hot Jupiters) as a function of the maximum considered orbital period.  The AD test conducted on the Q6 catalog is shown for reference (green dotted).  There is a significant excess of single planet systems with short orbital periods which indicates both that such planets can survive and that there is a population of small planets that is somewhat analogous to the hot Jupiters.\label{shortperhist}}
\end{figure}

Figure \ref{shortperhist} shows the PDF of orbital periods for single and multiplanet systems out to a maximum orbital period of 20 days.  One can see a sizeable excess of single planet systems with orbital periods less than a few days.  We used both the Kolmogorov-Smirnov (KS) and the Anderson-Darling (AD) test to determine the statistical significance of the differences in these samples.  The two tests were repeated for several sub-samples of differing maximum orbital periods---from one to 20 days in one-day increments.  The bottom panel in Figure \ref{shortperhist} shows the results of these tests for the different sub-samples.

With a maximum orbital period of only two days the $p$-values from both tests are of order 0.01---already quite small given the relatively few multiplanet systems.  By three days the $p$-values are below $10^{-4}$.  Also shown in Figure \ref{shortperhist}, for illustrative purposes, are the results of the AD test over the same range using data from the Q6 catalog \citep{Batalha:2013} which shows a less significant hint of excess single planets \citep{Latham:2011}.  Using the Q12 catalog the statistical significance of the difference grows by a large amount.  We note that if one extends this analysis to longer orbital periods the $p$-values rise until the maximum period is $\sim 50$ days, after which the $p$-values drop and settle to a constant value of order $10^{-6}$.

Observations of the single planet systems show that a larger fraction of them are astrophysical false positives \citep{Santerne:2012} compared to multiplanet systems \citep{Lissauer:2012}.  However, the \citet{Santerne:2012} results apply primarily to giant planets (excluded from our study) and the false positive rate that they observed was greatest among candidates with orbital periods $\gtrsim 5$ days---beyond the primary regime considered here.  Moreover, with smaller planet candidates, viable astrophysical false positive scenarios become increasingly scarce \citep{Fressin:2013}---indicating that the number of false positives among the single planet systems considered here should be less than the 35\% found by \citet{Santerne:2012}.  Regardless, if astrophysical false positives were randomly distributed among the single planet systems, repeating the tests with a 50\% false positive rate produces only a small change in the AD and KS test results when the maximum orbital period is beyond $\gtrsim 2.5$ days.

These statistical tests illustrate a couple of points.  First, a fairly large number of small planets can indeed survive in orbits less than a few days provided that they are in single-planet systems.  In multiplanet systems, the innermost planet can also survive, but the outer planets in the systems must be located at an increasingly larger period ratio from the inner planets.  Second, that there is a statistically significant excess of short-period, single planet systems---perhaps a small-planet counterpart to the 3-day ``pile-up'' of hot Jupiter planets.  If this is the case, then it is not obvious that these hot, smaller planets arrived at their locations through the same processes that drove the hot Jupiter migration since some of these processes have important mass dependence (e.g., planet-planet scattering \citet{Ford:2008b}).

\section{Projected Distribution Analyses}
\label{misc}

If the observed dearth of planet pairs in close proximity to the host stars is due to some late-stage differential migration of the orbits, then one might expect a feature in Figure \ref{pratios} such as a cluster of points at short orbital periods but with large period ratios or a build-up of planets roughly parallel to the boundary described by Equation \ref{boundaryfunc}---though such a feature would likely be smoothed out due to the distribution of planet and stellar masses.  Figure \ref{pratios} shows no significant clustering of points in the upper-left portion of the plot (detection efficiency precludes a similar group of points in the upper right).  However, the Q12 catalog does show a larger number of points in that region which, with additional scrutiny and more data of various types, may prove important.

\begin{figure}[!ht]
\includegraphics[width=0.45\textwidth]{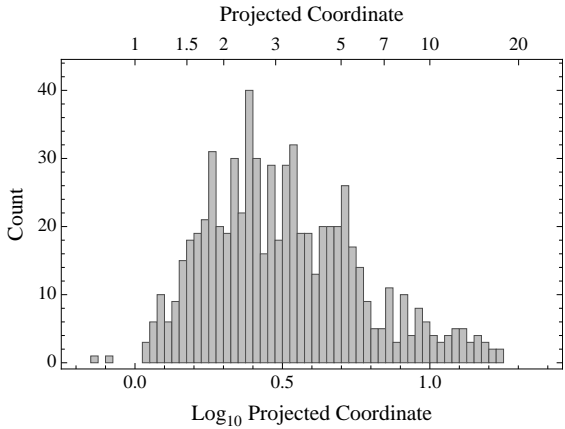}
\caption{The distribution of points from Figure \ref{pratios} projected onto a line perpendicular to Equation \ref{boundaryfunc}.  If the cause of the missing systems resulted in the inner planet drifting inward, then we might expect a large peak in this distribution just to the right of zero.\label{projectedhist}}
\end{figure}

To see if a clustering of points parallel to Equation \ref{boundaryfunc} is seen we produce a histogram of the data from Figure \ref{pratios} projected onto a line perpendicular to Equation \ref{boundaryfunc} and given by
\begin{equation}
\prat \sim P_1^{3/2}.
\end{equation}
This histogram is shown in Figure \ref{projectedhist}.  This figure does not show a significant excess of planets near the boundary given by Equation \ref{boundaryfunc}.

If there is a physical cause of the absent planet pairs, then there is something particular about the Kepler-42 system.  Kepler-42 (KOI-961) is a tiny planetary system orbiting a small star of mass $0.13 M_\odot$ and radius of $0.17 R_\odot$ \citep{Muirhead:2012}.  The effective temperature and metallicity for Kepler-42 are $T_\textrm{eff} = 3100$K and [Fe/H] = $-0.5$.  This places Kepler-42 among the smallest of known, planet-hosting main sequence stars (indeed it is among the smallest of all main sequence stars).  The stellar parameters for this system in the Kepler Input Catalog (KIC) are quite far from those obtained from spectroscopic observations---an expected difference since the KIC strategy was not optimized to identify late M stars, but one to consider when studying the stars hosting some \kepler\ systems.

\citet{Muirhead:2012} estimates that there are only a few dozen mid to late M dwarfs being observed by \kepler .  Thus, among the planetary systems shown in Figure \ref{pratios} it is possible, but somewhat unlikely that there are other planet-hosting \kepler\ targets similar to Kepler-42.  More data on planets orbiting M dwarf stars may yield other systems that can be included in the analysis performed here.  Regardless of whether or not planetary systems hosted by such small stars satisfy Equation \ref{boundaryfunc}, the results should prove interesting.

\section{Discussion}

This study showed that the lack of close-proximity multiplanet systems with orbital periods less than a few days appears to be quite significant.  However, the cause of this feature in the planet population is not known.  Since these planetary systems are so close to their host stars, there are several physical processes that may play a role in shaping the system architectures.  Tides, disk interactions, winds, General Relativity, and magnetic fields are some obvious suspects.  Perhaps the differences in the stellar tidal forces for planet pairs in such small orbits tends to drive the inner planets into the host star (or the outer planet away from the star---depleting the lower left of Figure \ref{pratios}).  Perhaps differential torques on the planets by the planet-forming disk pulls their orbits apart, or the swift migration of the outer planet might drive the inner planet into the star.  It may be that the cause of the depleted region is also the source of the excess single-planet systems with small orbital periods (since the mechanism that produces them may not be the same as the hot Jupiters).  Whatever the explanation, it must not predict a significant cluster of points in the upper-left of Figure \ref{pratios} nor a significant excess of points near the boundary in the projected distribution Figure \ref{projectedhist}.

We note that the fact that Kepler-42 is the only system that violates Equation \ref{boundaryfunc} may be a simple consequence of the low mass of the host star---the forces that produce the feature may only effective when the planets are closer.  It's curious departure from the rest of the systems may prove the key to identifying the cause of the observations.  Moreover, the fact that the points for Kepler-42 in Figure \ref{pratios} lie on a line nearly parallel to Equation \ref{boundaryfunc} may not be coincidental.  Indeed, many of the three-planet systems with short orbital periods produce lines that are roughly parallel to that equation.  However, the bias against detecting very long period planets with companions at large period ratios (the upper right portion of Figure \ref{pratios}) precludes definitive statistical statements on this issue at this time.

Additional data on these \kepler\ targets and theoretical investigations into the cause of this anomalous lack will give important insights into the history and nature of these small-sized planetary systems.  Improved mass estimates of the stars near the boundary are likely to be important.  With improved masses, the orbital distance can be determined and from that insights may be gained into the cause of the absent systems.  Table \ref{targettable} is the list of some system properties of the KOIs shown in the figures sorted by the distance from Equation \ref{boundaryfunc} of the nearest planet pair in each system.  This table could be the basis for an observation campaign to better characterize the planet hosting stars.  In addition to characterizing known systems, the discovery of more short-period multiplanet systems (particularly with NASA's TESS mission) will be useful.

\begin{table*}
\begin{minipage}{6.0in}
\begin{center}
\caption{Stellar information from the NASA Exoplanet Archive (Q8) for each system sorted by the minimum distance of any planet pair from the line given in Equation \ref{boundaryfunc} (measured perpendicular to that line).\label{targettable}}
\begin{tabular}{llllllll}
KID & KOI & $T_\textrm{eff}$ (K) & $\log g$ & $R_\star$ & Kp & $P_\textrm{min}$ (d) & Min Log Distance \\ \hline
8561063 & K00961 & 4188 & 4.562 & 0.681 & 15.92 & 0.4532875 & -0.134227525 \\
9006186 & K02169 & 5447 & 4.42 & 0.93 & 12.404 & 2.192589 & 0.033651729 \\
7595157 & K00568 & 5390 & 4.61 & 0.77 & 14.14 & 2.359002 & 0.037548792 \\
6685609 & K00665 & 6080 & 4.37 & 1.1 & 13.182 & 1.611881 & 0.048465628 \\
5809890 & K01050 & 5088 & 4.55 & 0.76 & 13.999 & 1.269095 & 0.050608858 \\
6871071 & K02220 & 6022 & 4.48 & 0.97 & 14.686 & 1.897809 & 0.052831842 \\
6962977 & K01364 & 5447 & 4.47 & 0.88 & 15.956 & 2.580788 & 0.060530303 \\
5972334 & K00191 & 5696 & 4.52 & 0.88 & 14.991 & 0.7085982 & 0.061067929 \\
10397751 & K02859 & 5464 & 4.47 & 0.88 & 13.851 & 2.005396 & 0.06374295 \\
7376983 & K01358 & 4601 & 4.66 & 0.67 & 15.505 & 2.34587 & 0.06546141 \\
\end{tabular}
\end{center}
\end{minipage}
\end{table*}

\section*{Acknowledgements}
Funding for the \kepler\ mission is provided by NASA's Science Mission Directorate.  We thank the entire Kepler team for the many years of work that is proving so successful and we deeply regret the fact that its extraordinary life was cut short by such a small wheel.  J.H.S. acknowledges support by NASA under grant NNX08AR04G issued through the Kepler Participating Scientist Program.  This research has made use of the NASA Exoplanet Archive, which is operated by the California Institute of Technology, under contract with the National Aeronautics and Space Administration under the Exoplanet Exploration Program.

\bibliographystyle{apj}

\clearpage

\end{document}